\documentclass[traditabstract]{aa}
\usepackage{graphicx}
\usepackage[]{natbib}
\usepackage{enumerate}
\def\lunits{$\rm erg\,s^{-1}$~}
\def\funits{$\rm erg\,cm^{-2}\,s^{-1}$~}
\def\cunits{$\rm cm^{-2}~$}
\def\sax{{\it BeppoSAX~}}
\def\xmm{{\it XMM-Newton~}}
\def\chandra{{\it Chandra~}} 
\def\lxl6{{$\rm L_X/L_{6\,\mu m}~$}}
\def\micron{{$\rm \mu m~$}}

\begin{document}
\title{On the Lx-L$_{\rm 6\,\mu m}$  ratio as a diagnostic for Compton-thick AGN}

%   \subtitle{ }

  \titlerunning{X-ray to $\rm 6\,\mu m$ luminosity ratio}
    \authorrunning{I. Georgantopoulos et al.}

   \author{I. Georgantopoulos\inst{1,2},
                    E. Rovilos \inst{1},
                    A. Akylas\inst{2},
                    A. Comastri \inst{1},
                    P. Ranalli \inst{1},
                    C. Vignali \inst{1},
                    I. Balestra \inst{3},
                    R. Gilli \inst{1}, 
                    N. Cappelluti \inst{1}
                     }

   \offprints{I. Georgantopoulos, \email{ioannis.georgantopoulos@oabo.inaf.it}}

   \institute{INAF-Osservatorio Astronomico di Bologna, Via Ranzani 1, 40127, Italy \\
              \and 
              Institute of Astronomy \& Astrophysics,
              National Observatory of Athens, 
              Palaia Penteli, 15236, Athens, Greece \\
              \and
              Max-Planck Institut f\"{u}r Extraterrestrische Physik,  Giessenbachstrasse 2, 85748 Garching, Germany \\
                 }

   \date{Received ; accepted }

\abstract{As the mid-IR luminosity represents a good isotropic
proxy of the AGN power, a low X-ray to mid-IR luminosity ratio is often claimed to be a reliable
indicator for selecting Compton-thick AGN. We assess the efficiency of the
 X-ray to mid-IR luminosity ratio diagnostic by examining the 12\,\micron
{\it IRAS} AGN sample (intrinsic $\rm L_{2-10 keV}>10^{42}$\,\lunits) for which high
signal-to-noise \xmm observations have been recently become available. We find
that the vast majority (ten out of eleven) of the AGN that have been classified as
Compton-thick on the basis the X-ray spectroscopy by Brightman \& Nandra
present a low \lxl6 luminosity ratio, i.e. lower than a few percent of the
average AGN ratio which is typical of reflection-dominated Compton-thick
sources. We note that at low \lxl6 ratios we also find a comparable number
of AGN, most of which are heavily absorbed but not Compton-thick. This implies
that although most Compton-thick AGN present low \lxl6 ratios, at least in the
local, Universe, the opposite is not necessarily true. Next, we extend our
analysis to higher redshifts. We perform the same analysis in the Chandra Deep
Field South where excellent quality \chandra (4\,Ms) and \xmm (3\,Ms) X-ray
spectra are available. We derive accurate X-ray luminosities for  \chandra
sources using X-ray spectral fits, as well as 6\micron luminosities from
spectral energy distribution fits. We find eight AGN (intrinsic
$\rm L_{2-10\,keV}>10^{42}$\,\lunits) with low \lxl6 ratios in total, after
excluding one source where the 6\,\micron emission primarily comes from star-formation.
One of these sources has been already demonstrated to host a Compton-thick
nucleus, while for another one at a redshift of $z=1.22$ we argue it is most
likely Compton-thick on the basis of its combined \chandra and \xmm spectrum. 
 In agreement with the low redshift sample, we find a large number of
non Compton-thick `contaminants' with low X-ray to mid-IR luminosity ratios. 
 The above suggest that a low
\lxl6 ratio alone cannot ascertain the presence of a Compton-thick AGN, albeit
the majority of low \lxl6 AGN are heavily obscured. More interestingly,
the two most reliable Compton-thick AGN in the high redshift Universe have high
\lxl6 ratios, showing that this method cannot provide complete Compton-thick
AGN samples.
       \keywords {X-rays: general; X-rays: diffuse background;
X-rays: galaxies; Infrared: galaxies}}
   \maketitle
%
%________________________________________________________________

\section{Introduction} 

The deep X-ray Universe has been probed at unparalleled depth thanks to the
\chandra mission. The deepest \chandra X-ray surveys revealed an AGN sky
density above 10.000$\rm deg^{-2}$ \citep{Alexander2003,Luo2008,Xue2011}, probing 
flux limit of $f_{2-10\,keV}\approx 1 \times 10^{-16}$\,\funits
\citep[for a review see][]{Brandt2005}. In contrast, the optical surveys for
QSOs reach sky densities of about few hundred per square degree at a magnitude
limit of $B=22$mag \citep[e.g.][]{Wolf2003}. Optical spectroscopic methods, which
detect AGN using the [OIII] ($\rm \lambda 5007 \AA$) or the [NeV] lines 
($\rm \lambda 3427 \AA$), \citep{Hao2005,Bongiorno2010,Gilli2010}, manage to
recover a large fraction of low-luminosity AGN \citep{Georgantopoulos2010} but these 
are applicable only at a limited redshift range, because these lines
are redshifted outside the optical band at high redshifts. Therefore, hard (2-10\,keV) X-ray
surveys have provided so far the most powerful method for the detection of AGN
and thus have provided the most detailed picture of the accretion history of
the Universe \citep[e.g.][]{Ueda2003, LaFranca2005, Aird2010}.

However, even the efficient hard X-ray surveys may be missing a substantial
fraction of the most heavily obscured sources, the Compton-thick AGN, which
have column densities $\rm >10^{24}\,cm^{-2}$. At these high column densities, 
the attenuation of X-rays is mainly due to Compton-scattering rather
than photoelectric absorption. Although there are only a few dozen of 
Compton-thick AGN identified in the local Universe,
\citep[see][for a review]{Comastri2004}, there is concrete evidence for the
presence of a much higher number. The peak of the X-ray
background at 20-30\,keV \citep[e.g.][]{Frontera2007,Churazov2007,Moretti2009}
can be reproduced by invoking a significant number of Compton-thick
sources at moderate redshifts. However, the exact density of Compton-thick
sources required by X-ray background synthesis models still remains an open
issue \citep*{Gilli2007,Sazonov2008,Treister2009}. Additional evidence for the
presence of a numerous Compton-thick population comes from the directly
measured space density of black holes in the local Universe
\citep[see][]{Soltan1982}. It is found that this space density is a factor of
1.5-2 higher than that predicted from the X-ray luminosity function
\citep{Marconi2004,Merloni2008}, although the exact number depends on the
assumed efficiency in the conversion of gravitational energy to radiation.

The {\it INTEGRAL} and {\it SWIFT} missions explored the X-ray sky at energies
above 10\,keV with the aim to provide the most unbiased samples of
Compton-thick AGN in the local Universe. Owing to the limited imaging
capabilities of these missions (coded-mask detectors), the flux limit probed is
very high ($\sim 10^{-11}$\,\funits), allowing only the detection of AGN at very low
redshifts. These surveys did not detect large numbers of Compton-thick sources
\citep[e.g.][]{Ajello2008,Tueller2008,Paltani2008,Winter2009,Burlon2011}. The
fraction of Compton-thick AGN in these surveys does not exceed a few percent of
the total AGN population. In contrast, optical and mid-IR 
surveys yield Compton-thick AGN fractions which range between 10 and 20\%
\citep{Akylas2009,Brightman2011}. As \citet{Burlon2011} point out, it is
possible that even these ultra-hard surveys are biased against the most heavily
obscured  ($>2\times 10^{24}$\,\cunits) Compton-thick
sources.

At higher redshifts, a number of efforts have been made to identify 
Compton-thick AGN by means of either X-ray hardness ratios \citep{Brunner2008,Gilli2011} 
 or X-ray spectroscopy
\citep{Tozzi2006,Georgantopoulos2009,Comastri2011,Feruglio2011} in
the deepest \chandra and \xmm observations. In particular, \citet{Comastri2011},
provided the most direct X-ray spectroscopic evidence yet for the presence of
Compton-thick nuclei at high redshift, reliably identifying two Compton-thick
AGN at z=1.536 and z=3.700.

The mid-IR wavelengths have attracted much attention for providing an
alternative way to detect heavily obscured systems. This is because the
absorbed radiation by circumnuclear dust is re-emitted in the IR part of the
spectrum \citep*{Soifer2008}, rendering heavily obscured AGN copious mid-IR
emitters. In such systems, the 2-10\,keV X-ray emission can be diminished 
by almost two orders of magnitude \citep[e.g.][]{Matt2004}, while at the same
time the isotropic mid-IR emission remains largely unattenuated.
\citet{Daddi2007,Fiore2008,Georgantopoulos2008,Treister2009b,Eckart2010,Donley2010}
propose that a fraction of infrared excess, $\rm 24\,\mu m$-bright sources
which are undetected in X-rays are associated with Compton-thick AGN. Their
main argument is based again on the fact that the X-ray to IR luminosity
ratio is very low, suggestive of significant amounts of absorption
\citep[but see][]{Georgakakis2010,Georgantopoulos2011}.

The detection of a low \lxl6 luminosity ratio has recently been used as the
main instrument for the detection of faint Compton-thick AGN which remain
undetected in X-rays \citep[e.g.][]{Alexander2008,Goulding2011}. This is
because the $\rm 6\,\mu m$ luminosity is an excellent proxy of the AGN power,
as it should be dominated by very hot dust which is heated by the AGN
\citep[e.g.][]{Lutz2004,Maiolino2007}. At these wavelengths the contribution
by the stellar-light and colder dust heated by young stars should be minimal.
\citet{Yaqoob2011} criticise the above results claiming that the \lxl6
ratio is not a robust indicator of Compton-thick obscuration. They perform
Monte-Carlo X-ray spectral simulations of heavily obscured AGN. They caution that
Compton-thick AGN may have the same X-ray to mid-IR luminosity ratio with heavily obscured AGN
which have a column density of a few times $10^{23}$\,\cunits, depending on the exact value
of the covering fraction and photon index. 

In this paper, we attempt to assess the efficiency of the \lxl6 ratio method,
i.e. to estimate the percentage of Compton-thick AGN among low \lxl6 sources,
but also to evaluate whether there are any bona-fide Compton-thick AGN with
substantially high \lxl6 ratios. First, we are using local AGN, the IRAS
12\,\micron sample of \citet*{Rush1993}, to assess the reliability and
effectiveness of the \lxl6 ratio selection. For this sample there are excellent
quality X-ray spectroscopic observations available \citep{Brightman2011}, and
therefore we know a priori which objects are Compton-thick. Second, we are
selecting candidate Compton-thick AGN among the low \lxl6 AGN in the CDFS. We
are presenting a detailed X-ray spectral analysis of these sources, using the
most sensitive X-ray data ever obtained, the 4\,Ms \chandra data, combined with
the 3\,Ms \xmm data. Our aim is to detect unambiguous signs of Compton-thick
obscuration, such as direct detection of a large column density, or a flat
spectral index, or a high equivalent-width Fe\,K$\alpha$ line. We adopt
$\rm H_o=75\,km\,s^{-1}\,Mpc^{-1}$, $\rm\Omega_{M}=0.3$ and $\Omega_\Lambda=0.7$
throughout the paper.

\section{Data}

\subsection{The IRAS 12\micron sample}

\subsubsection{XMM data} 

The extended IRAS 12\,\micron galaxy survey \citep{Rush1993} is a sample of 893
MIR selected local galaxies, which contains a high fraction of AGN (over 10\%
according to optical spectroscopy). The sample is taken from the {\it IRAS} Faint
Source Catalogue, version 2 (FSC-2) and imposes a flux limit of 0.22\,Jy,
including only sources with a rising flux density from 12 to 60 microns (in
order to minimise the contamination by stars) and with a Galactic latitude of
$|b|>25^\circ$. There are \xmm observations available for 126 galaxies from the
12\,\micron sample.

\subsubsection{Mid-IR data}

For the local sources, we used photometry from the NASA Extragalactic Database (NED). The most commonly used
datasets are the 2MASS survey \citep{Skrutskie2006} for the near-IR, and
{\it IRAS}, {\it ISO}, and {\it Spitzer} catalogues for the near-IR mid-IR and far-IR.
We used only broad-band photometry datapoints, and avoided measurements made
with apertures smaller than the apparent sizes of the sources.
The number of data-points used range from seven (IRAS07599+6508) 
 up to 33 (Mrk231).

\subsection{CDFS}

\subsubsection{Chandra}

The CDFS 4\,Ms observations consist of 53 pointings obtained in the years 2000
(1\,Ms), 2007 (1\,Ms) and 2010 (2\,Ms). The analysis of the first 1\,Ms data is
presented in \citet{Giacconi2002} and \citet{Alexander2003}, while the analysis
of the 23 observations obtained up to 2007 is presented in \citet{Luo2008}. 462
X-ray sources have been detected in the 2\,Ms data by \citet{Luo2008}. The
average aim point is $\alpha=03^h32^m28^s.8$,
$\delta=-27^\circ48^{\prime}23^{\prime\prime}$ (J2000). The observations cover an
area of about 440\,arcmin$^2$. The present CDFS 4\,Ms survey reaches a
sensitivity limit of $\sim0.7\times 10^{-16}$\,\funits and
$1\times 10^{-17}$\,\funits in the hard (2-8\,keV) and soft (0.5-2\,keV) band
respectively \citep{Xue2011}. The Galactic column density towards the CDFS is
$0.9\times 10^{20}$\,\cunits \citep{Dickey1990}.

\subsubsection{XMM-Newton}

The CDFS area was surveyed with \xmm in several different epochs spread over
almost nine years. The data presented in this paper are obtained combining the
observations awarded to our project in AO7 and AO8 \citep{Comastri2011}. These
observations are performed in four different epochs between July 2008 and March
2010, while the archival data were obtained in the period July 2001 - January
2002. The total exposure, after cleaning from background flares, is
$\approx$2.82\,Ms for the two MOS and $\approx$2.45\,Ms for the PN detectors.
An extended and detailed description of the full data set including data
analysis and reduction and the X-ray catalogue will be published in Ranalli et
al. (in preparation).

\subsubsection{Mid-IR}

The central regions of the CDFS have been observed in the mid-IR by the
{\it Spitzer} mission as part of the Great Observatory Origin Deep Survey
(GOODS). These observations cover areas of about $10 \times 16.5\rm\,arcmin^2$
in both fields using the IRAC (3.6, 4.5, 5.8 and 8.0$\, \rm \mu m$) and the
MIPS (24 and 70\,$\rm \mu m$) instruments on board {\it Spitzer}. 
 We construct a uniform dataset of infrared fluxes by matching  the 2\,Ms X-ray catalogue
of \citet{Luo2008} with the IRAC catalogue from the SIMPLE survey
\citep{Damen2011}, which covers the total area of the CDFS. 
 We find  counterparts for 446 out of the 462 sources. We also use the FIDEL survey of the
CDFS \citep[see][]{Magnelli2009} to get photometric information on longer
wavelengths ($\rm 24\,\mu m$ and $\rm 70\,\mu m$).
 There are 161 sources with mid-IR data in the four IRAC bands, 210 sources 
  with five bands (IRAC + MIPS 24 \micron) and finally 69 sources with mid-IR data 
   in six bands (IRAC+  MIPS, 24 \micron and 70 \micron). For the remaining six sources we have only 
    information in less than four IRAC bands. For these sources we do not derive spectral energy distribution 
    (SED) fits.  

\section{Analysis}

\subsection{12 \micron sample}

For the local sample we are using the \xmm spectroscopic information presented 
in \citet{Brightman2011}. These authors provided X-ray spectral fits with
particular emphasis on revealing the Compton-thick sources. There are 60 high
X-ray luminosity sources (i.e. with {\it intrinsic} $\rm L_X>10^{42}$\,\lunits)
among the 126 sources, with \xmm observations presented in
\citet{Brightman2011}. These 60 sources are classified as bona-fide AGN on the
basis of the high X-ray luminosity alone. The X-ray luminosity classification
is based on the fact that the most X-ray luminous star forming galaxies
\citep*[e.g. NGC\,3256;][]{Moran1999}, present a 2-10\,keV luminosity below
$10^{42}$\,\lunits, without showing evidence for AGN activity. Among these 60
sources, there are ten AGN which show X-ray spectral evidence for Compton-thick
obscuration according to \citet{Brightman2011}. That is they present in their
X-ray spectra either
a) high equivalent-width (EW) Fe$K_\alpha$ lines, or
b) flat spectral indices indicative of a reflection component, or
c) direct detections of an absorption turn-over with $\rm N_H>10^{24}$\,\cunits.

 To construct the \lxl6 diagram, we obtain the X-ray luminosities
(uncorrected for absorption) for the AGN in the 12\micron sample using the
\xmm fluxes from \citet{Brightman2011}. The 6\micron IR flux has been
estimated using SED fits (see
Sect.\,\ref{SEDs}). 

\subsection{CDFS}

\subsubsection{Chandra spectra}
For the CDFS sample we derive X-ray spectra for the sources in the
\citet{Luo2008} catalogue using the 4\,Ms data \citep{Xue2011}.
  
  We used the {\sc SPECEXTRACT} script in the {\sc CIAO} v4.2 software package to
extract the spectra of all 462 X-ray sources in our sample. The extraction
radius varies between 2 and 4 arcsec with increasing off-axis angle. At low
off-axis angles ($<$4\,arcmin) this area encircles 90\% of the light at an energy of
1.5\,keV. The same script extracts response and auxiliary files. The addition
of the spectral files was performed with the
{\sc FTOOL} task {\sc MATHPHA}. For the addition of the response and auxiliary files 
 we used the {\sc FTOOL}  {\sc ADDRMF}, and {\sc ADDARF} respectively,
  weighting according to the number of photons in each spectrum.

 We fit the spectra using an absorbed power-law model 
  {\sl PLCABS}  \citep{Yaqoob1997} which properly takes into account
Compton-scattering up to column densities of $\rm N_H\sim 10^{25}$ \cunits.  
We used the {\sc XSPEC} v12.5 software package for the spectral fits
\citep{Arnaud1996}.
We used the C-statistic technique \citep{Cash1979} specifically developed
to extract spectral information from data with a low signal-to-noise ratio. 
 We derive accurate X-ray
luminosities from the X-ray spectral fits. This is the most reliable method for
deriving X-ray luminosities. Instead, the derivation of the X-ray luminosity
through the X-ray count rate and an average photon-index can distort
substantially the \lxl6 relation.

\subsubsection{XMM-Newton spectra} 
For each individual \xmm orbit, source counts were collected from a circular region of 
12 arcsec radius, centered on the source position. 
Local background data were taken from nearby regions, separately for the PN, MOS1 and MOS2 detectors 
 where no source is found in the full exposure image.
  The area of the background region had  20 arcsec radius. 
  The spectral data from individual exposures were summed up for the source and background, respectively, 
  and the background subtraction was made assuming the common scaling factor for the source/background geometrical areas. 
   Both PN and MOS spectra are extracted in the 0.5-8 keV energy area. 
    MOS1 and MOS2 spectra are summed using the FTOOLS {\sl MATHPHA} task.
     Response and effective area files were computed by averaging the individual files 
     using the FTOOLS {\sl ADDRMF} and {\it ADDARF} tasks.   
     
\subsection{Spectral energy distributions}
\label{SEDs}

\begin{figure*}
\begin{center}
\includegraphics[width=9.0cm]{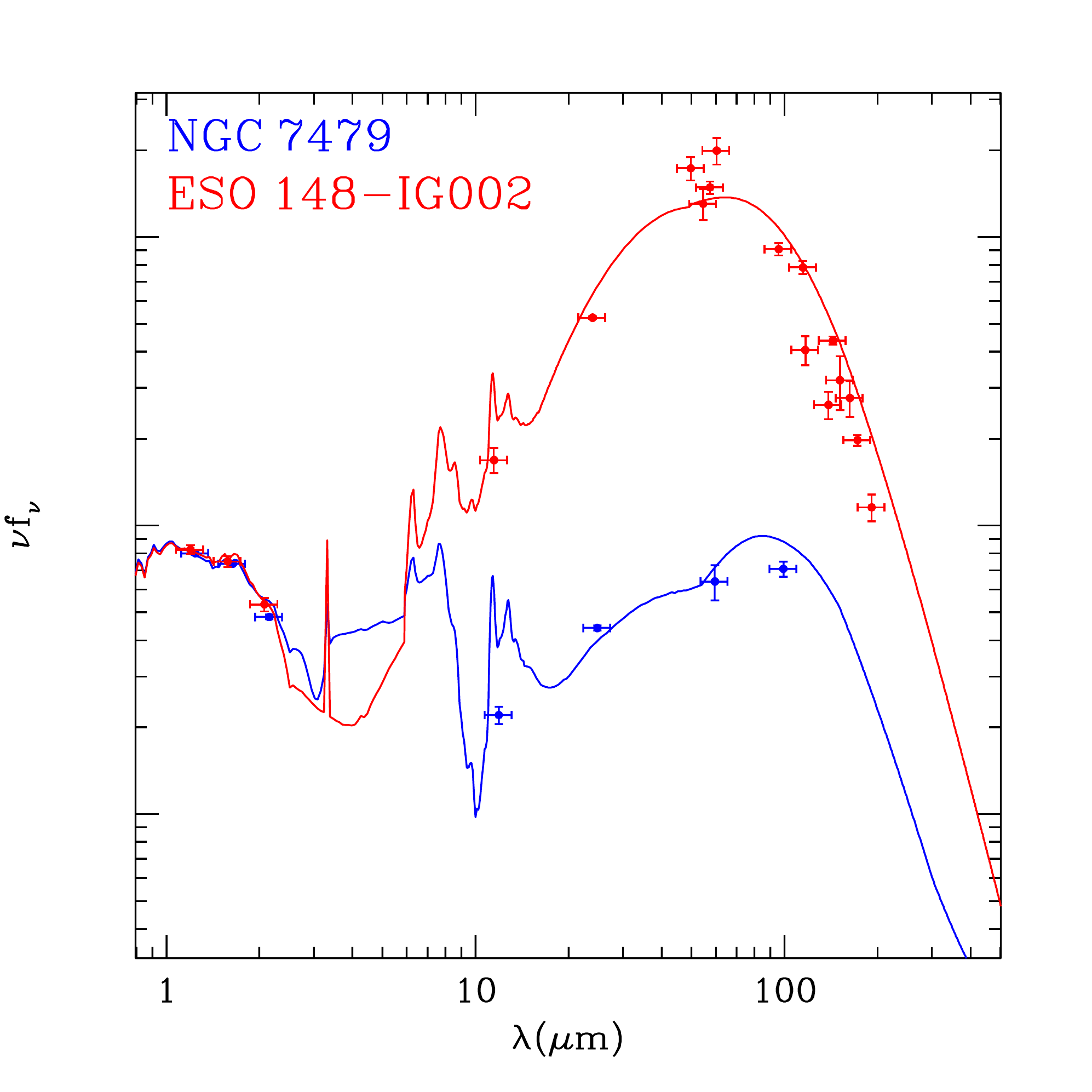}
\includegraphics[width=9.0cm]{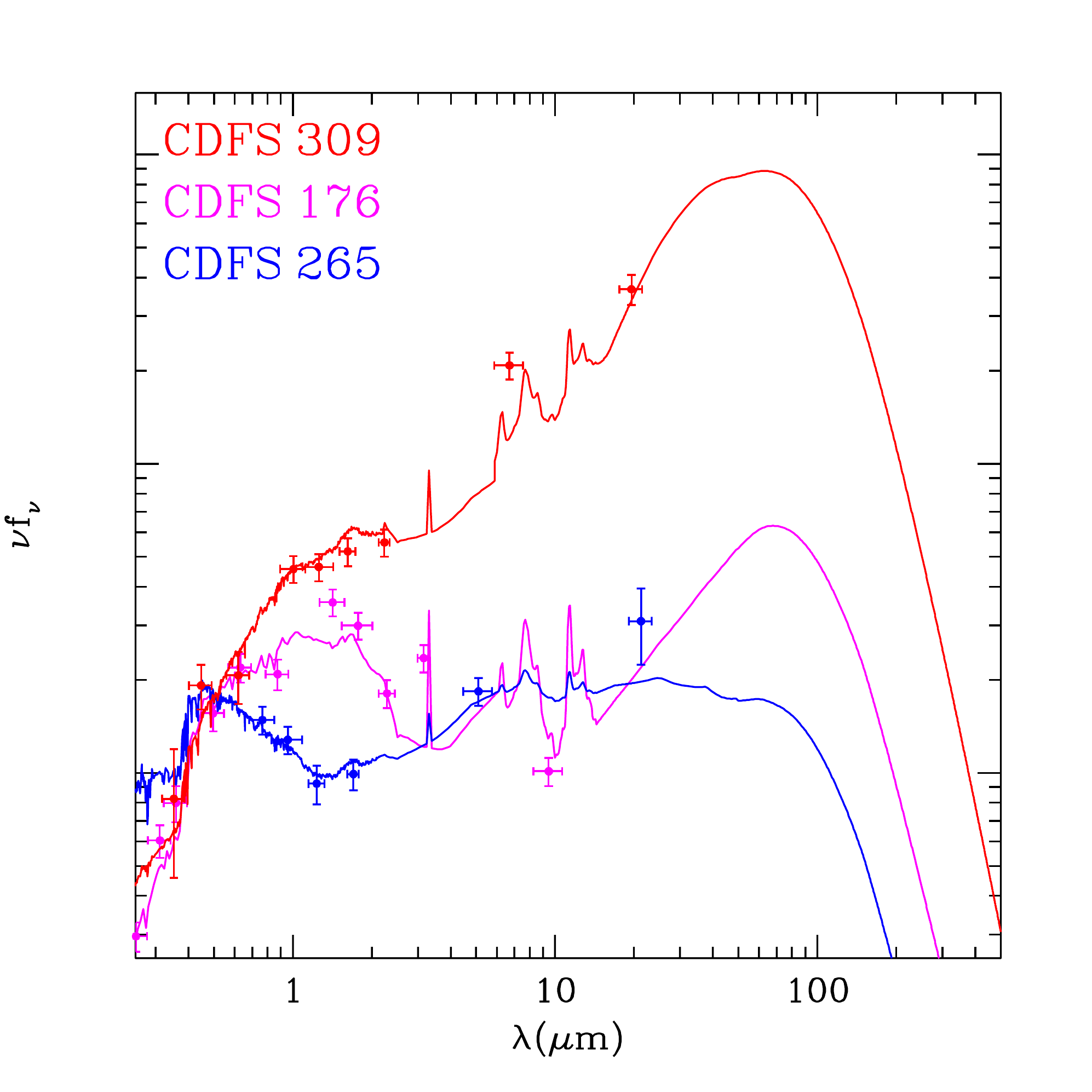} 
\caption{Left panel: the SED of the Compton-thick sources NGC7479 and ESO148-IG002 
 from the 12$\mu m$ sample. Right panel: the SED of the two Compton-thick sources 
  in \citet{Comastri2011} in comparison with that of the candidate Compton-thick source 
   CDFS-309 in \citet{Feruglio2011}. The SED have been 
   normalised to have the same flux at 0.5$\rm \mu m$.}
\label{SED}
\end{center}
\end{figure*}

We constructed SED to obtain an idea on the
dominant powering mechanism (AGN or star-formation) in the mid-IR part of the
spectrum. The SEDs are also used in order to obtain an accurate estimate of the
$\rm 6\mu m$ infrared luminosities. We used a $\chi^2$ minimisation technique to fit the
optimum combinations of host galaxy and pure AGN SED templates to the
broad-band spectra of our objects. The starburst templates we use, are those from the
SWIRE template library of \citet{Polletta2007}, which is a compilation of
observed SED of nearby galaxies (M82, NGC6090, Arp220, IRAS20547, IRAS22491).
In addition, we use the templates  of \citet{Chary2001}. 
The AGN templates we use come from \citet*{Silva2004}, who combine nuclear SEDs 
of Seyferts with a range of absorption columns. We also construct a sample 
of AGN templates using the type-1 QSO SED from \citet{Richards2006} and
applying dust absorption following \citet*{Rosenthal2000}, in order to account
for the 9.7\,\micron absorption feature.
In Fig. \ref{SED}, we show an example  SED of Compton-thick sources in both the high 
 and low-redshift Universe.

\section{Results} 

\subsection{The IRAS 12\micron sample}

\begin{figure*}
\begin{center}
\includegraphics[width=12.0cm]{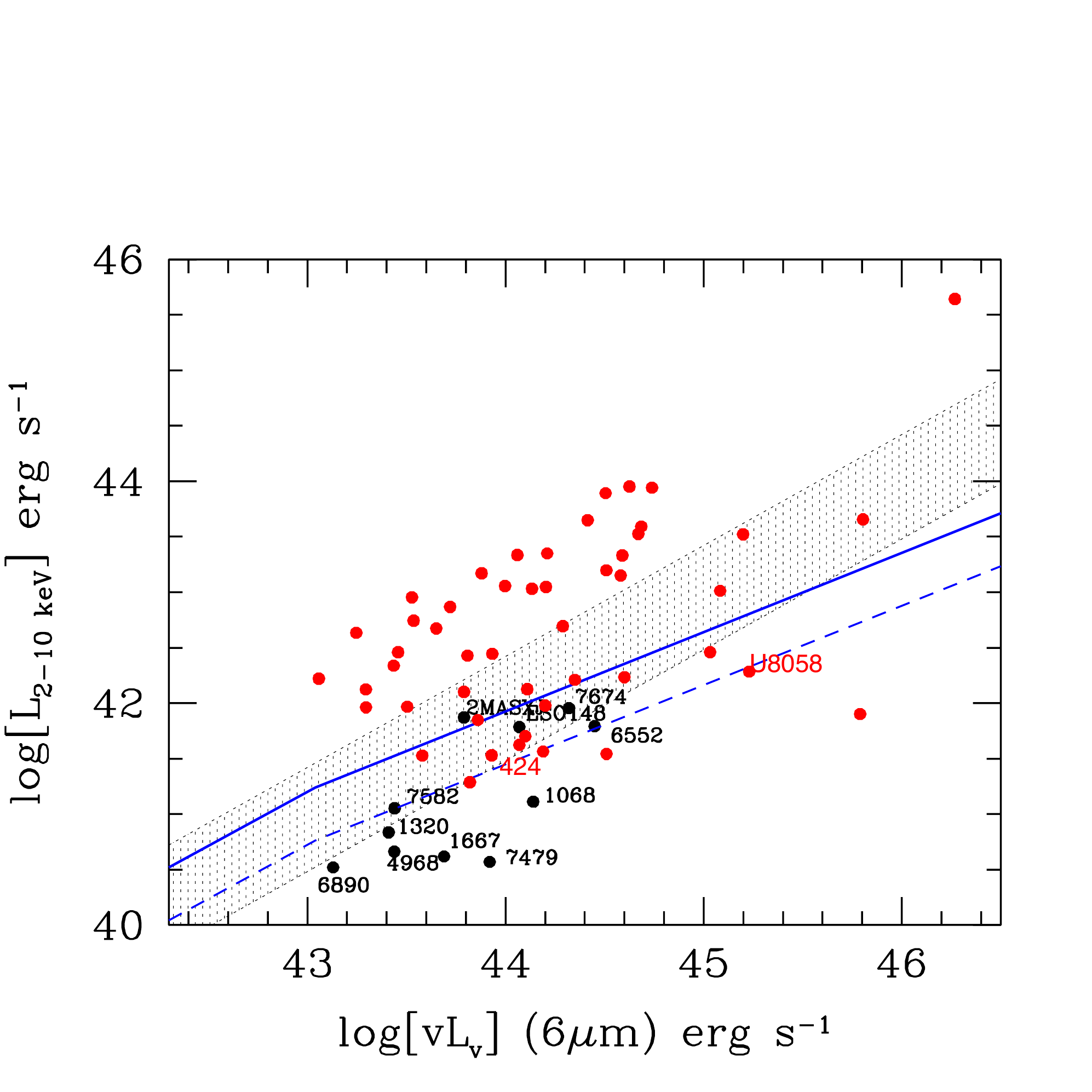} 
\caption{Rest-frame  $\rm L_X$ vs. $\rm L_{6\,\mu m}$
         luminosity diagram for the unambiguous AGN i.e. those with {\it intrinsic} X-ray 2-10 keV luminosity
          $L_X>10^{42}$ \lunits.  The X-ray  luminosity is uncorrected 
          for absorption while the mid-IR luminosity includes  both the torus 
           and the star-formation component. The black and the red circles correspond to the
         Compton-thick and Compton-thin sources in the AGN ($\rm L_X>10^{42}$\,\lunits) sample of
         \citet{Brightman2011}. The shaded area 
         denotes the Compton-thick regime based on the average \lxl6 relation (corrected for obscuration) 
         found for local AGN by \citet{Lutz2004} scaled down assuming a 3\% fraction of reflected 
          emission from the backside of the obscuring screen in the 2-10 keV band.  The dispersion in this relation 
          is  $\approx$0.7dex for type-2 AGN. The blue lines show
         the luminosity-dependent average relation for Compton-thick AGN based
         on the COSMOS AGN  assuming a 3\% (solid-line) and a 1\% (dashed-line) fraction of reflected emission. 
         Only one Compton-thick
         source (2MASX\,J15504152) lies above the nominal Compton-thick
         regime as defined on the basis of the \citet{Fiore2009} average relation assuming a 3\% reflection.}
   \label{brightman-lxl6}
\end{center}
\end{figure*}

We present the \lxl6 diagram for the nearby sources in
Fig. \ref{brightman-lxl6}.  The X-ray luminosity is uncorrected for obscuration while 
 the 6$\rm \mu m$ luminosity includes both the torus and the star-formation 
  component (see discussion). The lines denote various Compton-thick AGN
 areas. These have been derived in the following fashion. We adopt the average
\lxl6 luminosity dependent AGN relation of \citet{Fiore2009} derived in the
COSMOS field. We then scale down this relation by a factor of  30. This is 
 an approximation to the average
reflected luminosity relative to the intrinsic emission from the backside of the torus in the 2-10\,keV band in
reflection-dominated AGN \citep[e.g.][]{Comastri2004}. Alternatively, 3\% is
the relation between the absorbed and unabsorbed 2-10\,keV luminosity for a
column density of $1.5\times10^{24}$\,\cunits and a power-law spectrum of
$\Gamma=1.8$. 
 We show an alternative scenario, where the relation of \citet{Fiore2009} 
 is scaled for only 1\% reflected emission \citep{Vignali2004}.  
We also give the \lxl6 relation derived by
\citet{Lutz2004} in the local Universe, together with its associated $1\sigma$ uncertainty 
 scaled down again by 3\% (dashed area).
Throughout the paper, we use only the \citet{Fiore2009} 3\% reflection limit for assessing the
number of Compton-thick sources. We see that most (ten out of eleven) of the sources
classified as Compton-thick on the basis of X-ray spectroscopy lie below the
Compton-thick lines. However, we see that many more sources (twelve) which are not 
Compton-thick, according to the \xmm X-ray spectroscopic diagnostics, would be
classified as candidate Compton-thick AGN on the basis of the low \lxl6 ratio. All these
 are presented in Table \ref{brightman}. It is not unlikely that some of
the low \lxl6 sources are indeed Compton-thick sources, but \xmm failed to 
identify them owing to the limited photon statistics. At least two of them
NGC424, \citep{Iwasawa2001,Burlon2011}  and UGC8058 (Mrk231), \citet{Braito2004}, are  
Compton-thick AGN according to {\it SWIFT} and {\it BeppoSAX} spectroscopy. 

%%%%%%%%%%%%%%%%%%%%%%%%%%%%%%%%%%%%%%%%%%%%%%%%%%%%%%

\begin{table*}
\centering
\caption{Candidate Compton-thick sources in the local sample according to the \lxl6 ratio.} 
\label{brightman} 
\begin{tabular}{cccccccc}
\hline\hline 
Name              & $z$    & $\rm L_X$ & $\rm L_{6\,\mu m}$ & fraction & CT  & $\rm N_H$     & EW         \\
(1)               & (2)    & (3)       & (4)            & (5)      & (6) & (7)       & (8)              \\
\hline
NGC\,17           & 0.0196 & 41.29     & 43.82          & 0        & -   & $47^{+30.4}_{-21.2}$ & 0.440   \\
NGC\,424          & 0.0118 & 41.60     & 43.93          & 1        & -   & $23.6^{+44.7}_{-15.8}$ & $<$0.024 \\
NGC\,1068         & 0.0380 & 41.11     & 44.14          & 0.74     & b   & -       &  -               \\
NGC\,1194         & 0.0136 & 41.53     & 43.58          & 0.91     & -   & $67.8^{+18.3}_{-16.4}$ & $<$0.059 \\
NGC\,1320         & 0.0890 & 40.83     & 43.41          & 0.83     & a   & $204^{+91}_{-99}$    & 0.082   \\
NGC\,1667         & 0.0152 & 40.62     & 43.69          & 0        & a   & $271^{+5730}_{-251}$  & 0.391  \\
F\,07599+6508 & 0.1488 & 41.90     & 45.79          & 0.96     & -   & -             & 0.027         \\
UGC\,5101         & 0.0394 & 41.70     & 44.10          & 0        & -   & $49.6^{+25.4}_{-18.2}$ & 0.229 \\ 
Mrk231        & 0.0422 & 42.28     & 45.23          & 0.95     & -   & $7.1^{+1.8}_{-1.4}$   &  0.011 \\
NGC\,4968         & 0.0099 & 40.66     & 43.44          & 0.92     & a   & $300^{+2230}_{-123}$  & 0.172  \\
NGC\,5256         & 0.0279 & 41.62     & 44.07          & 0        & -   & $17.5^{+5.6}_{-3.8}$  & 0.608  \\
Mrk\,273          & 0.0378 & 41.98     & 44.20          & 0        & -   & $59.7^{+17.1}_{12.8}$  & 0.192 \\
Mrk463         & 0.0504 & 42.46     & 45.03          & 0.98     & -   & $38.7^{+9.2}_{-5.7}$  &  $<$0.008 \\
Mrk848           & 0.0402 & 41.54     & 44.51          & 0.79     & -   & $85.9^{+2030}_{-62.2}$ & - \\
\dag2MASX\,J15504152  & 0.0303 & 41.87     & 43.79          & 0.48     & c   & $0.36^{+0.54}_{-0.28}$ & - \\
NGC\,6552         & 0.0265 & 41.79     & 44.55          & 0.95     & b   & -          &        -      \\
F\,19254-7245      & 0.0617 & 42.23     & 44.60          & 0.88     & -   & $38.1^{+39.2}_{-21.7}$ & 0.064 \\
NGC\,6890         & 0.0081 & 40.52     & 43.13          & 0.24     & c   & $0.11^{+0.20}_{-0.09}$ & 0.237 \\
NGC\,7479         & 0.0079 & 40.57     & 43.92          & 0.70     & a   & $201^{+493}_{-122}$   & - \\
NGC\,7582         & 0.0053 & 41.05     & 43.44          & 0        & c   & 0.04   & -                \\
NGC\,7674         & 0.0289 & 41.95     & 44.32          & 0.58     & c   & -       & 0.132                \\
ESO\,286-IG019    & 0.0430 & 41.56     & 44.19          & 0.85     & -   & $48.7^{+258}_{-35.2}$  & - \\
ESO\,148-IG002    & 0.0446 & 41.78     & 44.07          & 0        & c   & -     & -                  \\
\hline \hline 
\end{tabular}
\begin{list}{}{}
\item The columns are:
(1) Name;
(2) Spectroscopic redshift;
(3) Logarithm of luminosity in the 2-10\,keV band uncorrected for absorption in
    units of \lunits;
(4) Logarithm of 6\,\micron luminosity in units of \lunits;
(5) Fraction of the AGN contribution to the $\rm 6\,\mu m$ luminosity;
(6) Compton-thick classification from \citet{Brightman2011} based on:
    a: direct measurement of the column density;
    b: large EW of Fe\,K$\alpha$ line;
    c: large reflection component;
(7) $\rm N_H$ column density in units of $\rm 10^{22}$ \cunits from \citet{Brightman2011};
(8) Equivalent-width of the 6.2\micron PAH in units of \micron from \citet{Wu2009};
\dag: Source not in the low-\lxl6 sample.
\end{list}
\end{table*} 
    
%%%%%%%%%%%%%%%%%%%%%%%%%%%%%%%%%%%%%%%%%%%%%%%%%%%%%%%%%%%%%%%%%%

%%%%%%%%%%%%%%%%%%%%%%%%%%%%%%%%%%%%%%%%%%%%%%%%%%%%

\begin{figure*}
\begin{center}
\includegraphics[width=12.0cm]{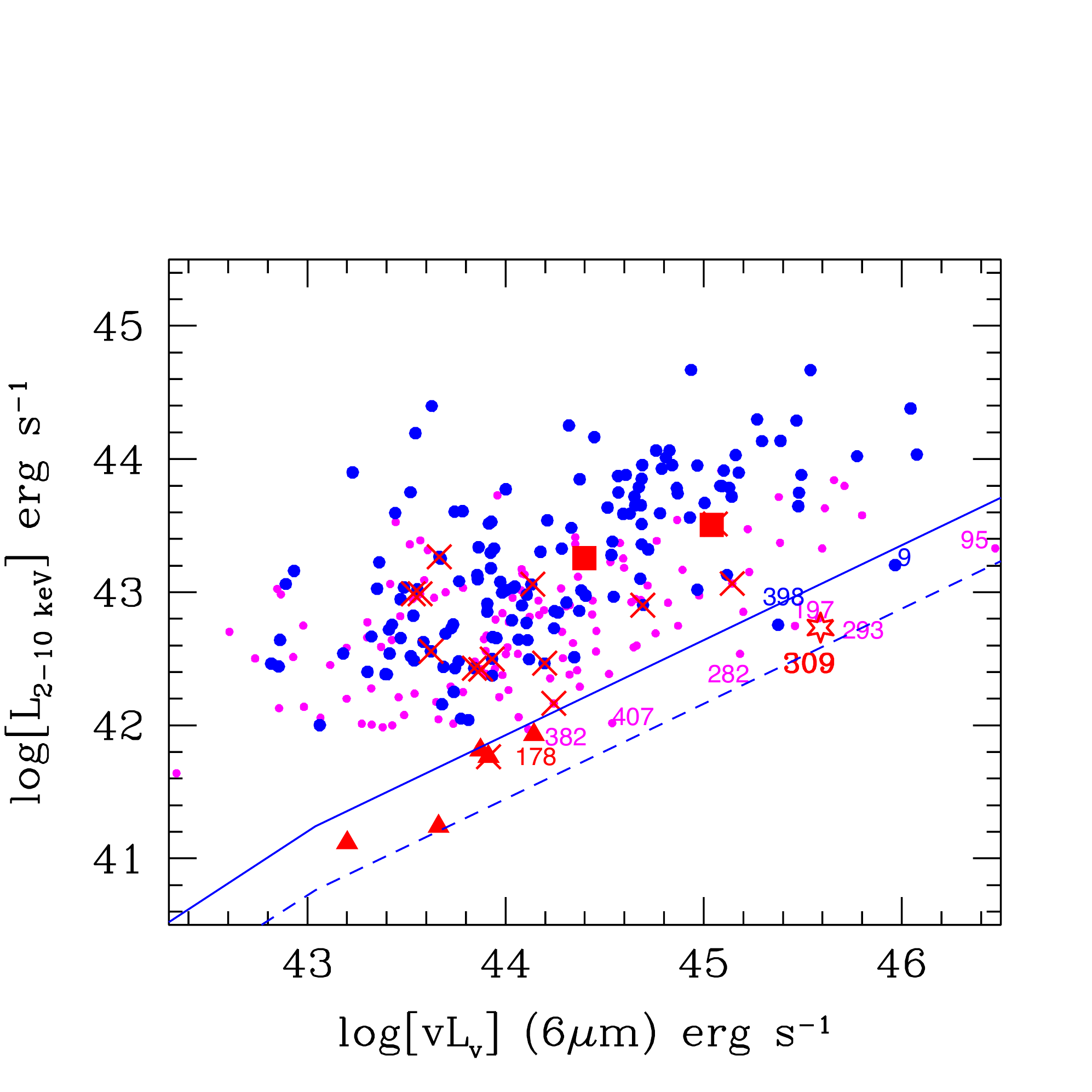} 
\caption{Rest-frame (uncorrected for absorption) $\rm L_X$ vs. $\rm L_{6\,\mu m}$
         luminosity diagram. The large filled circles (blue) correspond to the
         bright sources ($\rm f_{2-10}>10^{-15}$ \funits), while the small circles
         (magenta) correspond to the fainter sources. 
         The red squares correspond to the two bona-fide
         Compton-thick AGN from the 3\,Ms XMM observations \citep{Comastri2011},
         while the red-star denotes source CDFS-309, proposed as a
         Compton-thick AGN by \citet{Feruglio2011}.
         The crosses refer to  
         the 20 candidate  Compton-thick sources of \citet{Tozzi2006} in the 1\,Ms
         CDFS observations.  These  
          include both the reflected  as well as the transmission-dominated sources;
           two sources which are not detected in \citet{Luo2008} and two more sources 
            which are detected in less than four IRAC bands have been excluded.  The red triangles
         correspond to the five low-luminosity sources ($\rm L_X<10^{42}$\,\lunits) with
         flat spectrum, given in Table \ref{lowlum}. The solid line 
         denotes the \lxl6 average Compton-thick AGN relation; this has been 
         derived from the average AGN relation of \citet{Fiore2009}, scaled down
         to the expected Compton-thick AGN X-ray luminosity, assuming a 3\%
         fraction of torus reflected emission relative to the intrinsic
         2-10\,keV luminosity. The dashed line  shows the 
         luminosity-dependent Compton-thick average \lxl6 relation for an 1\%
         torus reflected emission.}
   \label{cdfs-lxl6}
\end{center}
\end{figure*}

\begin{table*}
\caption{Chandra spectral fits and 6\,\micron luminosities of the low \lxl6
          sources in the CDFS}
\centering
\begin{tabular}{cccccccccccccc} 
\hline \hline 
ID  & $z$   &    $\rm N_H$ &  $\Gamma$ &  C-stat& $\rm f_{2-10}$ & $\rm L_X$ & $\rm L^{tot}_{6\,\mu m} $ & fraction & $\rm N_H$             & C-stat  & $\Gamma$               & C-stat    & counts \\ 
(1) & (2)   & (3)            & (4)       & (5)                       & (6)                        & (7)                   & (8)     & (9)                    & (10)      & (11) & (12) & (13) & (14)\\  
\hline
9   & 1.615 &  $10.8^{+4.6}_{-5.0}$ & $2.53^{+0.73}_{-0.84}$ & 438 & 1.5            & 1.8       & 11.77  & 0.92  & $8.7^{+4}_{-3}$       & 439 & $0.88^{+0.15}_{-0.25}$ & 456 & 130  \\          
95  & 5.73$\dagger$  & $83.4^{+8.7}_{-7.6}$ &  $1.80^{+1.10}_{-0.85}$ & 505 & 1.0  & 2.7  & 12.96  & 1.0 & $94_{-38}^{+29}$      & 505 & $0.94^{+0.26}_{-0.35}$ & 508 & 137  \\
197 & 3.64$\dagger$  & $126^{+365}_{-125}$   &   $2.65^{+5.3}_{-2.4}$ & 474 & 0.46 & 0.59      & 11.74   & 0.97  & $86^{+39}_{-50}$      & 437 & $0.55^{+0.50}_{-0.65}$ & 472 &37  \\
282 & 2.223 & -  &   -  & - & 0.14  & 0.38      & 11.95  & 0.87  & $<$1.2                & 477 & $2.98^{+1.14}_{-0.96}$ & 472  & 24 \\
293 & 3.10  & -   &    -  & -  & 0.15  & 0.65      & 11.75   & 0.96   & $16.2^{+21}_{-15}$    & 452 & $1.09^{+0.75}_{-0.53}$ & 455.5 & 22\\
309 & 2.579 & -  &    -   & -  & 0.35  & 1.5       & 11.61   & 0.81   & $30^{+24}_{-14}$      & 497 & $0.60^{+0.57}_{-0.57}$ & 498  & 52  \\
382 & 0.664 & $26.8^{+29.5}_{-16.0}$  &   $2.27^{+0.80}_{-0.65}$  & 487 & 0.63   & 0.1       & 10.75   & 0.1    & $1.4^{+0.77}_{-0.66}$ & 488 & $1.46^{+0.21}_{-0.44}$ & 497  & 179\\
398 & 1.222 & $3.2_{-3.0}^{+8.0}$  &   $2.02_{-0.9}^{+0.47}$   & 483 & 1.50   & 0.54      & 11.77   & 0.88  & $3.6^{+2.8}_{-1.8}$   & 470 & $0.81^{+0.30}_{-0.32}$ & 465  &94 \\
407 & 1.367 & $<$3.9 &    $1.46_{-0.73}^{+1.32}$  & 464 &  0.13  & 0.12      & 10.83   & 0.35  & $<$3.5                & 464 & $1.48^{+1.26}_{-0.72}$ & 464   & 48\\ 
\hline \hline 
\end{tabular}
\begin{list}{}{}
\item The columns are:
(1) ID in the catalogue of \citet{Luo2008};
(2) Redshift from \citet{Luo2010}; two and three decimal digits denote
    photometric and spectroscopic redshift respectively; $\dagger$ source not detected in the BVR bands \citet{Cardamone2010} or K band
     \citet{Taylor2009}  and thus the photometric redshift may be uncertain. \citet{Rafferty2011} gives a photometric redshift of 
      z=2.30 and z=3.02 for the sources CDFS-197 an CDFS-293 respectively; 
 (3) Column density in units $10^{22}$\,\cunits, (4) Photon-index; 
   and (5) C-statistic value for 509 degrees of freedom for both $\Gamma$ and $\rm N_H$ free;
(6) X-ray flux in the 2-10\,keV band in units of $10^{-15}$\,\funits, as
    estimated from the X-ray spectroscopy;
(7) X-ray luminosity in the 2-10\,keV band (uncorrected for X-ray absorption) in units of
    $10^{43}$\,\lunits;
(8) Logarithm of total 6\,\micron luminosity (star-formation and torus) in
    units of solar luminosity; 
(9) Fraction of torus contribution relative to total luminosity at 6\,$\mu m$,
    according to SED decomposition;
(10) Column density in units $10^{22}$\,\cunits for photon index fixed to
    $\Gamma=1.8$;
    (11) C-statistic value for 510 degrees of freedom;
    (12) Photon-index for column density $\rm N_H=0$ \cunits; 
    (13) C-statistic value for 510 degrees of freedom;
    (14) Net counts (background subtracted) in the 0.5-8 keV band; 
    All errors quoted refer to the 90\% confidence level.    
\end{list}
\label{cdfs}
\end{table*} 

\begin{table*}
\caption{Chandra and XMM X-ray spectral fits of sources CDFS-9 and CDFS-398}
\centering
\begin{tabular}{ccccccc} 
\hline \hline 
model       & mission       & $\rm N_H$            & $\Gamma$              & EW                     & C-stat    \\ 
(1)         & (2)           & (3)                  & (4)                    & (5)  & (6)             \\  
\hline
\multicolumn{6}{c}{CDFS-9} \\
\hline 
PL          & XMM           & $6.4^{+3.1}_{-2.2}$  & $1.43^{+0.34}_{-0.28}$  & -                      & 2892/2497 \\
PL  + Line & XMM        & $5.9^{+2.9}_{-2.2}$  & $1.41^{+0.33}_{-0.28}$ &   $0.29\pm0.21$ & 2886/2496 \\
PL          & Chandra + XMM & $8.2^{+2.8}_{-2.6}$ & $1.64^{+0.29}_{-0.29}$  &  -                      & 3334/3008 \\
PL + Line   & Chandra + XMM & $7.5^{+2.8}_{-2.4}$ & $1.58^{+0.20}_{-0.29}$ & $0.18^{+0.10}_{-0.10}$ & 3316/3007 \\ 
\hline
\multicolumn{6}{c}{CDFS-398} \\
\hline 
PL         & XMM           & $<0.68$                 & $0.67^{+0.26}_{-0.29}$   & -                      & 2947/2499 \\
PL + Line & XMM        &  $<0.62$    &  $0.68^{+0.30}_{-0.27}$   &  $0.71\pm0.49$     & 2940/2498 \\
PL         & Chandra + XMM & $<$0.74  & $0.86^{+0.26}_{-0.26}$    & -                      & 3405/3008 \\
PL + Line  & Chandra + XMM & $<$0.33 & $0.78^{+0.24}_{-0.21}$ & $0.60\pm0.13$  & 3395/3007 \\
PEX + Line & Chandra + XMM & -                      & $2.68^{+0.3}_{-0.3}$ & $0.18^{+0.06}_{-0.06}$ & 3407/3010 \\
\hline \hline 
\end{tabular}
\begin{list}{}{}
\item The columns are: 
(1) Model; 
(2) Data used in the fits;
(3) Column density in units of $10^{22}$\,\cunits; 
(4) Photon Index;
(5) EW (rest-fame) of the FeK$\alpha$ line; 
(6) C-statistic value divided by degrees of freedom; All errors correspond to
    the 90\% confidence level.
 \end{list}
\label{xmm}
\end{table*}

\begin{table*}
\caption{Low \lxl6 and low intrinsic X-ray luminosity sources with flat spectra.}
\centering
\begin{tabular}{ccccccccc} 
\hline \hline
ID  & $z$   & $\rm f_{2-10}$ & $\rm L_X$ & $\rm L^{tot}_{6\,\mu m}$ & $\Gamma$               & C-stat & $\rm N_H$           & C-stat \\ 
(1) & (2)   & (3)            & (4)       & (5)                      & (6)                    & (7)    & (8)                 & (9)    \\  
\hline
96  & 0.310 & 0.76           & 41.23     & 10.07                    & $0.83^{+0.43}_{-0.41}$ & 503    & $<$0.53             & 512    \\
159 & 0.664 & 0.78           & 41.81     & 10.28                    & $0.39_{-0.71}^{+0.51}$ & 539    & $<$0.24             & 549    \\
166 & 0.340 & 0.54           & 41.12     &  9.61                    & $0.33_{-0.77}^{+0.66}$ & 438    & $4.4^{+4.7}_{-1.9}$ & 440    \\
178 & 0.735 & 1.22           & 41.76     & 10.32                    & $-0.81^{+1.4}_{-2.0}$  & 452    & $158^{+83}_{-74}$   & 449    \\ 
208 & 0.738 & 0.72           & 41.92     & 10.55                    & $0.62^{+0.52}_{-0.52}$ & 488    & $<$0.75             & 496    \\
\hline \hline 
\end{tabular}
\begin{list}{}{}
\item The columns are:
(1) Name in the catalogue of \citet{Luo2008};
(2) Spectroscopic redshift from \citet{Luo2010}; 
(3) X-ray flux in the 2-10\,keV band in units of $10^{-15}$\,\funits, as
    estimated from the X-ray spectroscopy;
(4) Logarithm of the obscured X-ray luminosity in the 2-10\,keV band in units
    of \lunits;
(5) Logarithm of the total 6\,\micron luminosity (star-formation and torus) in\
    units of solar luminosity; 
(6) Photon index $\Gamma$ for column density fixed to $\rm N_H=0$, as derived
    from the \chandra spectral fits;
(7) Corresponding C-statistic value;     
(8) Column density in units of $\rm 10^{22}$ \cunits for photon index fixed to $\Gamma=1.8$; 
(9) Corresponding C-statistic value; All errors refer to the 90\% confidence level. 
\end{list}
\label{lowlum}
\end{table*}

\subsection{CDFS}

 We exclude all sources with an {\it intrinsic} X-ray luminosity of
$\rm L_X<10^{42}$\,\lunits, as many of these could be associated with normal galaxies. 
283 sources have luminosities above this level. 
The \lxl6 diagram is shown in Fig. \ref{cdfs-lxl6}. There are nine sources which would be
characterised as Compton-thick on the basis of the low \lxl6 ratio (see Table
\ref{cdfs}). In Table \ref{cdfs} we present the \chandra spectral fit results.  
 Throughout the paper, we use the 
 \citet{Luo2008} identification numbers. Four
low-\lxl6 sources could be classified as Compton-thick: CDFS-309, CDFS-95,
CDFS-197, and CDFS-398. CDFS-309  has been
discussed in detail in \citet{Feruglio2011} and was found to present a
Compton-thick source on the basis of an Fe\,K$\alpha$ line with a high EW.
 Due to the limited photon-statistics we cannot tell whether the source 
  is flat $\Gamma\approx0.6$ or obscured with $\rm N_H\sim 3\times 10^{23}$ \cunits. 
   However, the addition of an FeK$\alpha$ line improves the statistic by $\Delta C \approx20$ 
   yielding an EW of 3.6$\pm 1.8$ keV. 
CDFS-95 and CDFS-197 have column densities somewhat below the Compton-thick
limit ($\sim 9 \times 10^{23}$\,\cunits), but are nevertheless consistent with
being Compton-thick within the errors. However, for both sources there are 
 only photometric redshifts available and therefore the column density estimates are uncertain. 

\begin{figure}
\begin{center}
\includegraphics[width=7.0cm]{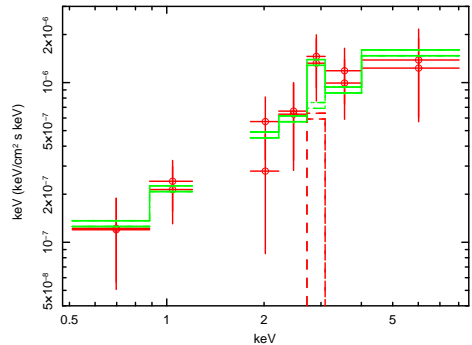} 
\caption{The XMM-Newton MOS X-ray spectrum of CDFS-398,
        together with the best-fit power-law model. The binning is such that
         there are about 15 net counts per MOS1 or MOS2  bin. The source counts amount to 
          30\% of the total (source+background) counts.}
   \label{xspec398}
\end{center}
\end{figure}

Only two low-\lxl6 sources have fluxes $\rm f_{2-10 keV} > 10^{-15}$\,\funits.
These sources have been also detected by \xmm, allowing a spectral fit with
both the photon-index and the column density considered as free parameters.
These are CDFS-9 and CDFS-398. In Table \ref{xmm} we present the spectral fit
results for these two sources. The spectrum of CDFS-9 is highly absorbed with a column density of
$\sim10^{23}$\,\cunits, but certainly it is not a Compton-thick source. From
Table \ref{xmm} we see that the \chandra spectrum of CDFS-398 is flat
($\Gamma\approx 0.8\pm 0.3$). Since the source is detected by \xmm, 
there are good photon statistics available allowing a more complex fit to the
spectrum. We consider two spectral models to fit its X-ray spectrum. 
(1) an absorbed power-law spectrum with the addition of an FeK$\alpha$ line,
    (plcabs+zga), and
(2) a reflection dominated model with an FeK$\alpha$ line, pexrav+zga.
The spectral fits to the combined \chandra and \xmm data are given in table
\ref{xmm}. This source displays a very flat spectrum $\Gamma < 1.2$ at the 90\%
confidence level. This flat spectrum is characteristic of reflection-dominated
Compton-thick sources \citep{Matt2004}. The {\it rest-frame} EW of the
Fe\,K$\alpha$ line is $0.68^{+0.66}_{-0.44}$\,keV, which is again consistent with
Compton-thick emission \citep{Fukazawa2011}. The X-ray spectrum is presented in
Fig. \ref{xspec398}. 

For clarity, we summarise the numbers of $\rm L_X/L_{6\mu m}$ 
 sources in the various subsamples given in his section in Table
\ref{numbers}.

\section{Discussion}

\subsection{The local sample}

In this work, we use the 60 bona-fide AGN, with luminosities
$\rm L_{2-10keV}>10^{42}$\,\lunits from the 12\,\micron {\it IRAS} sample.
 \xmm observations of these sources are presented in
\citet{Brightman2011} and \citet{Brightman2011b}. 22 of these AGN present low \lxl6 ratios. In particular,
their X-ray luminosity is lower than the 3\% of the average AGN
luminosity-dependent \lxl6 relation as derived by \citet{Fiore2009}. This is
assumed to be the fraction of the reflected 2-10\,keV emission from the
backside of the torus, relative to the intrinsic 2-10\,keV luminosity. Although
this value of reflected emission is often observed in the local Universe
\citep[e.g.][]{Comastri2004}, it should only be considered as a rough
approximation. For example, it is likely that in some cases it could be
even smaller \citep{Ueda2007, Comastri2010}. It is noteworthy that the \lxl6 method can
retrieve the vast majority (ten out of eleven) of the Compton-thick AGN found by
\citet{Brightman2011}. This ascertains that the empirical limit adopted above
is proven to be efficient for the selection of Compton-thick sources. The
X-ray spectroscopy reveals that ten out of the 22 low \lxl6 AGN are Compton-thick
AGN following the classification of \citet{Brightman2011}. Most of the
remaining twelve sources are heavily obscured having $\rm N_H>10^{23}$\,\cunits, but
these are not formally Compton-thick, at least according to the \xmm
spectroscopy. This suggests that the low \lxl6 ratio technique presents roughly
a 45\% success rate in the detection of Compton-thick sources in the local
Universe. Interestingly, the level of contamination is not extremely sensitive on
the exact value of the adopted \lxl6 upper limit. For example, from Fig.
\ref{brightman-lxl6}, we can easily estimate that by lowering the adopted
threshold e.g. by as much as a factor of three (i.e. adopting a 1\% reflection
fraction), we end up with eleven sources, six of which are classified as
Compton-thick by \citet{Brightman2011}.
The number of Compton-thick sources rises to seven taking into 
account UGC8058 which is a  Compton-thick source 
 according to \sax spectroscopy \citep{Braito2004} as noted above. 

One important parameter which is often ignored is the fraction of the 
star-formation contribution to the 6\,\micron emission. In most cases (15/23)
the torus component is the dominant contributor  to the 6\,\micron luminosity.
In the remaining cases the torus contribution to the 6\,\micron emission is
small i.e. $<$30\% (see Table \ref{brightman}). In the same table, we give the 
 equivalent-width (EW) of the PAH (Polycyclic Aromatic Hydrocarbon) 
  features at 6.2\micron as reported in \citet{Wu2009}. 
A high EW (usually above 0.3 \micron) is considered as an indication of a strong 
 star-forming component. We see that sources that have a small torus contribution,
  as derived from the SED fit,  generally display high EW.
   The only  marginal case is Mrk273 which displays no torus in its SED 
    while the EW of the 6.2\micron PAH feature is $\approx0.2$ \micron.
For the sources that display a dominant star-forming  contribution, one should be
cautious in using the \lxl6 ratio as a Compton-thick diagnostic, because the
above ratio represents only a lower-limit to the true value. In principle, in
these cases one could use instead the torus 6\micron luminosity. Nevertheless,
the degeneracies on the SED decomposition could play a crucial role, that is, 
in some cases the combination of different tori with star-forming models yield
equally good fits although the relative torus contribution changes. Such
degeneracies will be significantly overcome with the use of {\it Herschel} data
which will be able to provide tight constraints on the star-forming
contribution. When we reject the sources with low torus contribution, $<$30\%, we are
left with 14 low-\lxl6 sources (see Table 5). Out of these, six are Compton-thick according to
the \xmm spectroscopy (NGC1068, NGC1320, NGC4968, NGC6552, NGC7479, NGC7674).

\subsection{CDFS}

In most cases (seven out of nine) the torus component is the dominant contributor 
(over 80\%) to the
6\,\micron luminosity. In the case of CDFS-382, the SED
decomposition shows that  the torus contribution is negligible and therefore the
\lxl6 classification becomes problematic. Yet another source (CDFS-407)
presents a low (30\%) torus contribution to the 6\micron emission. However,
even taking into account the corrected 6\micron luminosity which comes from
the torus, we find that this source would be marginally classified as a low
\lxl6 ratio source.

In conclusion, in the CDFS, there are eight low \lxl6 AGN among the 283
luminous AGN ($\rm L_{2-10 keV}>10^{42}$\,\lunits), after exclusion of the CDFS-382 which has 
 zero torus contribution. At face value, i.e. assuming that all sources are
Compton-thick, this would correspond to a Compton-thick fraction of 3\%. In
agreement with the results in the local sample, not all the low \lxl6 AGN appear to be
associated with Compton-thick sources. On the basis of the flat spectra
(indicative of reflection dominated spectra), as well as the presence of
strong Fe\,K$\alpha$ lines, we estimate that at least two sources are
Compton-thick (CDFS-309 and CDFS-398). One of these (CDFS-309 at a redshift of
z$\approx$2.6) has been already reported as a Compton-thick source by 
\citet{Feruglio2011} while the other one (CDFS-398 at a redshift of z=1.222) is
reported here for the first time. Another two (CDFS-95, CDFS-197) are heavily
obscured ($\rm N_H\sim 10^{24}$\,\cunits) and thus consistent with being
Compton-thick. For these sources there are only photometric redshifts available 
and thus the column densities remain somewhat uncertain.

The fraction of Compton-thick AGN mentioned above (between two and four out of
283 sources) should be considered only as a lower limit. This is because there could be
many more Compton-thick AGN at lower luminosities. There are 85 additional
candidates with low \lxl6 luminosity ratio which have an intrinsic X-ray
luminosity $\rm L_X<10^{42}$\,\lunits in the 2-10 keV band. Many of these could
be associated with normal galaxies \citep*[e.g.][]{Georgakakis2003,Georgakakis2007,Tzanavaris2006}, as
well as low-luminosity AGN with no obscuration. For example,
\citet{Georgakakis2010} find that many low-luminosity AGN in the sample of
\citet{Terashima2002} present low X-ray to mid-IR luminosity ratios.
Nevertheless, there are five sources among them which show evidence for a flat
spectrum (see Table \ref{lowlum}). In principle these could be associated with 
heavily obscured nuclei. Unfortunately, as they have faint fluxes, it is
difficult to disentangle whether the hard spectra are caused by a moderate column
density, or a genuinely flat spectrum, which could be the trademark of a
reflection dominated Compton-thick AGN. There is no evidence for the presence
of an Fe\,K$\alpha$ line in these spectra.

\subsection{Comparison with previous results in the CDFS}

Interestingly, there are two bona-fide Compton-thick sources which do not
present low \lxl6 ratios. These are the sources CDFS-176 and CDFS-265 at a
spectroscopic redshift of $\rm z=1.536$ and $\rm z=3.700$ respectively. These sources have been
reported as Compton-thick by \citet{Norman2002}, \citet{Tozzi2006}, and
\citet*{Georgantopoulos2007}. Recently, \citet{Comastri2011} have
confirmed the presence of Compton-thick nuclei on the basis of both flat
spectra and high-EW Fe\,K$\alpha$ lines. The most likely explanation for a 
high X-ray to  6\,\micron luminosity ratio is the large intrinsic dispersion of this relation. 
This can be clearly seen in Fig. \ref{SED}. 
It appears that the Compton-thick AGN in \citet{Comastri2011}
 have a relatively low 6$\mu m$ luminosity compared to sources such as
   CDFS-309. Hence, only the last source lies below the Compton-thick line of Fig. 3. 
In any case, the existence of Compton-thick AGN which have high \lxl6
ratios certainly poses problems for the completeness of the \lxl6 method in
identifying Compton-thick AGN.

Additionally, \citet{Tozzi2006} have proposed a number of Compton-thick
candidates on the basis of flat spectra in the 1\,Ms observations of the CDFS
\citep[see also][]{Georgantopoulos2007}. All these candidates, with the
exception of CDFS-178, present high \lxl6 luminosity ratios (see Fig.
\ref{cdfs-lxl6}). This source, at a spectroscopic redshift of $z=0.735$, is
among our low X-ray luminosity sources which present a flat X-ray spectrum. 

\begin{table}
\caption{Numbers of low $\rm L_X/L_{6\mu m}$  AGN (intrinsic $\rm L_X>10^{42}$ \lunits) in the various samples,
 assuming the \citet{Fiore2009} relation for a 3\% reflection efficiency.}
\centering
\begin{tabular}{ccccccc} 
\hline \hline 
      & \multicolumn{3}{c}{Total} & \multicolumn{3}{c}{Torus$>$30\%} \\
      & all  & CT & non-CT & all  & CT & non-CT \\
\hline
Local & 22          & 10  & 12     & 14          & 6  &  8     \\
CDFS &  9          & 4  &  5     & 7(8)$^{\dagger}$       & 4  & 3(4) $^{\dagger}$   \\
\hline \hline 
\end{tabular}
\begin{list}{}{}
\item $^{\dagger}$ Source CDFS-407 has a low (30\%) torus contribution
      in the $\rm 6\,\mu m$ flux, but it would still be considered low \lxl6
      taking into account this contribution.
 \end{list}
\label{numbers}
\end{table}

\section{Summary and conclusions}

In this paper, we are assessing the efficiency of the low
$\rm L_X/L_{6\,\mu m}$ ratio technique for discovering Compton-thick AGN.
The principle behind this method is that in Compton-thick sources, the X-ray
rest-frame luminosity, is diminished by almost two orders of magnitude while the
6\,\micron luminosity is left unattenuated.

The advantages of the present work in comparison with previou applications
of the \lxl6 method are the following: 
\begin{itemize}
\item{We use a 'calibration' sample in the local Universe. 
      This consists of a sample with high-quality \xmm observations available
      \citep{Brightman2011b} and thus with a priori knowledge of the fraction of
      Compton-thick sources.}
\item{At higher redshift, we employ the most sensitive X-ray observations so far
      available, namely the 4\,Ms \chandra observations, combined with the
      3\,Ms \xmm observations in the CDFS \citep{Comastri2011}.}
\item{We derive the X-ray luminosities using proper X-ray spectral fits instead
      of the usually employed X-ray fluxes in combination with an average
      X-ray spectral index.}
\item{We use SED decomposition to ascertain that the nuclear
      contribution is not contaminated by significant amounts of star-forming
      emission.}
\end{itemize}

Our conclusions can be summarised as follows:
\begin{itemize}
\item{In the local sample the vast majority (ten out of eleven) of the Compton-thick AGN
      appear to have low \lxl6 ratios. However, the opposite is not true, i.e.
      there is a large number of low \lxl6 sources which are probably not
      Compton-thick. This would make the efficiency of the \lxl6 method for
      selecting Compton-thick sources 50\%.}
\item{In the CDFS we select eight low \lxl6 AGN with intrinsic luminosities
      $\rm L_X>10^{42}$\,\lunits (after exclusion of one source where the 
       6$\rm \mu m$ emission comes from star-formation).
       One of these (CDFS-309 at a redshift of
      $z\approx2.6$) is already shown to host a Compton-thick nucleus, while
      for another one (CDFS-398 at z=1.222) we argue here for the first time
      that it is Compton-thick.}
\item{A large fraction of the low \lxl6 CDFS sources cannot be confirmed as
      Compton-thick on the basis of the X-ray spectroscopy, but they are
      nevertheless highly obscured, having column densities above
      $10^{23}$\,\cunits. This suggests that the \lxl6 ratio cannot be used on
      its own to reliably classify sources as Compton-thick.}
\item{Finally, there at least two bona-fide Compton-thick sources from
      \citet{Comastri2011} which do not appear to present low \lxl6 ratios,
       casting further doubt on the validity of this method for selecting 
       Compton-thick sources.}  
\end{itemize}

\begin{acknowledgements}
IG and AC acknowledge the Marie Curie fellowship FP7-PEOPLE-IEF-2008 Prop.
235285. AC, RG and CV acknowledge receipt of ASI grants I/023/05/00 and I/88/06.
 NC acknowledges financial support from the Della Riccia foundation. 
The Chandra data used were taken from the Chandra Data Archive at the Chandra X-ray
Center. 
\end{acknowledgements}

\end{document}